\documentclass[aip,pop,reprint]{revtex4-1}
\usepackage[utf8]{inputenc} 
\usepackage[pdftex]{graphicx}
\usepackage{amsmath}
\usepackage{amssymb} 
\usepackage{epstopdf}

\graphicspath{{./figs/}}

\begin{document}

\title{Near-surface electron acceleration during intense laser-solid interaction in the grazing incidence regime}

\date{\today}
\author{D.~A.~Serebryakov}
\email{dms@appl.sci-nnov.ru}
\author{E.~N.~Nerush}
\author{I.~Yu.~Kostyukov}
\affiliation{Institute of Applied Physics of the Russian Academy of
Sciences, 46 Ulyanov St., Nizhny Novgorod 603950, Russia}

\begin{abstract}
When a relativistically intense $p$-polarized laser pulse is grazingly incident onto a planar solid-state target, a slightly superluminal field structure is formed near the target surface due to the incident and reflected waves superposition.
This field structure can both extract the electrons from the target and accelerate them.
It is theoretically shown that the acceleration is possible and stable for a wide range of electron initial conditions.
PIC simulations confirm that this mechanism can actually take place for realistic parameters.
As a result, the electron bunches with charge of tens of nC and GeV-level energy can be produced using a laser intensity $10^{21}$--$10^{22}$ W/cm${}^2$.
It is also shown that the presence of a preplasma can increase the acceleration rate, which becomes possible because of more efficient electron injection into the accelerating field structure.
\end{abstract}

\maketitle

\section{Introduction}

Laser-driven electron acceleration is nowadays a rapidly developing field.
It is attributed to the recent outstanding advances in manufacturing of lasers with extremely high peak power, which allow to generate ultrastrong electric fields (up to $\sim 10^{12}$ V/cm with currently available lasers).
As it is several orders of magnitude higher than achievable in conventional (microwave) electron accelerators, it is desirable to utilize these fields to accelerate the electrons (up to many GeV on a millimeter scale).
After that, high-energy electrons may be employed in, for example, bremsstrahlung-based X-ray/gamma-ray sources~[\onlinecite{Glinec2005, Cipiccia12, Li17gammaray}], all-laser-driven Compton X-ray/gamma-ray sources~[\onlinecite{Phuoc12, Powers14ComptonXrays, Banerjee15ComptonXrays}], neutron sources~[\onlinecite{Tsymbalov17}] or sources of coherent THz radiation~[\onlinecite{Leemans2003THz, Liao16THz}].

There are two most common approaches to the laser-driven acceleration problems: laser-plasma electron acceleration, where the interaction of the laser pulse with plasma may generate longitudinal electric fields which allow efficient acceleration (e.g. as in LWFA schemes [\onlinecite{Esarey09}]) and direct laser acceleration or vacuum laser acceleration [\onlinecite{Thevenet15}] where the laser field itself is used to accelerate an electron.
In order to allow electrons to gain energy over large distances, one usually needs to have a longitudinal electric field (so the electron which co-propagates with the laser pulse may be accelerated), but the electric field in a plane wave is fully transverse.
Different schemes have been proposed for creating a longitudinal field, such as utilizing highly focused laser pulses [\onlinecite{Salamin02}] or crossed laser pulses [\onlinecite{Haaland95, Esarey95, SalaminAPL2000}].
However, those schemes require an external electron injector which should be synchronized with the laser pulses on a femtosecond scale.
Another approach is based on the interaction of a laser pulse with dense planar targets~[\onlinecite{Baton2003, Naumova2004, Li2006, Chen2006, Zigler16, Andreev16, Brantov17}], structured targets~[\onlinecite{Fedeli16, Breuer14, Jiang16}] or waveguides~[\onlinecite{York2008, YiPukhovSrep16}], which allows production of sub-femtosecond multi-MeV electron bunches which usually have high charge because of the high electron density of the target.
Producing such sources of energetic electron bunches is of much interest among experimentalists over the last few years~[\onlinecite{Zigler16, Fedeli16, Breuer14, Culfa14}].

In the present work, we study analytically and numerically the configuration when a single $p$-polarized laser pulse is grazingly incident onto a planar solid-state target (see Fig.~\ref{fig:schematic}).
Due to incident and reflected waves superposition, a slightly superluminal field structure is formed near the target surface.
This field structure can both extract electrons from the target and accelerate them along the surface.
We theoretically show that the acceleration in the such field is possible on a wide range of electron initial conditions, so a significant fraction of the extracted electrons may be trapped and accelerated by the field, and high-charge electron bunches can be produced.
By means of particle-in-cell simulations, we demonstrate that the acceleration rate depends linearly on the laser field strength even at high normalized laser field amplitude $a_0$ (of the order of 50--100), although the interaction with the surface becomes highly nonlinear and the field configuration changes.
In the simulations, the generations of the electron bunches with energy up to GeVs and very high total charge (up to 17.5 nC) is observed.
We also demonstrate a positive effect of the presence of a preplasma on the bunch charge and electron  acceleration rate and show the existence of the optimal preplasma density, which could be important for experiments in this area.

\section{Near-surface electron acceleration}
\label{sec:acc-model}
Let us consider a plain $p$-polarized electromagnetic wave obliquely incident from vacuum onto an ideally reflecting surface. We introduce the coordinate system as shown in Fig.~\ref{fig:schematic}, with $x$-axis parallel to the surface, and $z$-axis normal to the polarization plane. 
If $\theta$ denotes the grazing angle (which equals $\pi/2-\psi$, where $\psi$ is the incidence angle), the incident (\textit{i}) and the reflected (\textit{r}) fields above the surface can be written as follows:
\begin{figure}
	\centering
	\includegraphics[width=0.7\columnwidth]{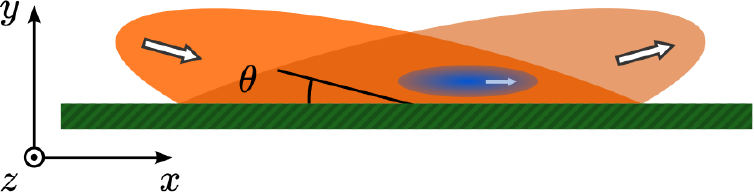}
	\caption{Grazing incidence of a laser pulse onto a reflective planar target. $\theta$ denotes a grazing angle, the electron bunch accelerated by the wave structure is shown in blue.}
	\label{fig:schematic}
\end{figure}
\begin{equation}\label{eq:incident}
\begin{aligned}
E_{x,i} &= E_0 \sin\theta \cos(kx\cos\theta - ky\sin\theta - \omega t + \phi_0),\\
E_{y,i} &= E_0 \cos\theta \cos(kx\cos\theta - ky\sin\theta - \omega t + \phi_0),\\
B_{z,i} &= B_0 \cos(kx\cos\theta - ky\sin\theta - \omega t + \phi_0),
\end{aligned}
\end{equation}
\begin{equation}\label{eq:reflected}
\begin{aligned}
E_{x,r} &= -E_0 \sin\theta \cos(kx\cos\theta + ky\sin\theta - \omega t + \phi_0),\\
E_{y,r} &= E_0 \cos\theta \cos(kx\cos\theta + ky\sin\theta - \omega t + \phi_0),\\
B_{z,r} &= B_0 \cos(kx\cos\theta + ky\sin\theta - \omega t + \phi_0),
\end{aligned}
\end{equation}
where $k = 2\pi/\lambda$ is a vacuum wavenumber, $E_0 = B_0$ is the field amplitude, $\phi_0$ is the initial phase.

So the superposition of incident and reflected fields yields 
\begin{equation}
\label{eq:oblique-fields}
\begin{aligned}
E_x &= 2 E_0 \sin\theta \sin(ky\sin\theta) \sin(kx\cos\theta-\omega t + \phi_0),\\
E_y &= 2 E_0 \cos\theta \cos(ky\sin\theta) \cos(kx\cos\theta-\omega t + \phi_0),\\
B_z &= 2 E_0 \cos(ky\sin\theta) \cos(kx\cos\theta-\omega t + \phi_0),
\end{aligned}
\end{equation}
which is a non-uniform wave which is running along $x$ with superluminal phase velocity $v_{ph} = c/\cos\theta$ and wavelength $\lambda_x = \lambda/\cos\theta$.
In the $y$-direction, the wave forms a standing structure with the spatial period $\Lambda = \lambda/\sin\theta$. 
One can verify that the field (\ref{eq:oblique-fields}) satisfies the boundary condition at $y=0$ ($E_x = 0$ and $B_y = 0$).
A snapshot of the fields~(\ref{eq:oblique-fields}) at $t=0$, $\phi_0 = 0$ and $\theta=15^\circ$ is shown schematically in Fig.~\ref{fig:N_gmax_px}(a), where arrows correspond to the electric field, and color shows the magnetic field.
It should be noted that the same field structure may be also formed by two crossed linear-polarized laser pulses~[\onlinecite{SalaminAPL2000}], so these formulas are applicable for that case as well.

\subsection{Maximum energy gain}

If we consider a test electron placed at $y = \Lambda/4$ at $t=0$, it turns out that it feels only $E_x$ component of the non-uniform wave ($E_y = B_z = 0$). 
If the electric field amplitude is of the order of $3\times10^{10}$ V/cm (which corresponds to dimensionless field amplitude $a_0 = eE_0/(mc\omega)=1$ for a typical laser wavelength $\lambda = 1~\mu m$) or higher, then the electron becomes relativistic during a fraction of the field period. 
So if the initial phase $\phi_0$ is properly chosen, it starts to move in the $x$-direction with speed $v_e$ which approaches the speed of light.
In turn, the wave propagates in the $x$-direction with the phase speed $v_{ph} = c/\cos\theta$ which is very close to the speed of light for $\theta \ll 1$. 
Therefore, the electron may experience the accelerating phase of the electric field for a long time, much longer than the field period $T = 2\pi/\omega$.

Such consideration is valid only for electrons that have exactly $y = \Lambda/4$ since this point may be an unstable equilibrium point in the $y$-direction, and it is not obvious whether the transverse instability will hinder the acceleration process. 
However, the instability may be suppressed for relativistic particles as the magnetic part and electric parts of the transverse Lorentz force almost compensate each other if $\theta\ll 1$ and $v\approx c$. 
This will be covered later in more detail.

One can also calculate the maximum energy gain in the above process based on the dephasing condition. 
Since the wave has phase speed $v_{ph} = c/\cos\theta$ which is slightly greater than $v_e \approx c$, it slowly overtakes the electron. 
After $N$ periods of the external field, the electron travels distance $L_e \approx cNT$ and a certain wave maximum travels $L_w = cNT/\cos\theta$. 
So the phase displacement of the electron equals
\begin{equation}
\Delta\varphi = 2\pi \frac{L_w - L_e}{\lambda_x} \approx \frac{2\pi}{\lambda_x} \left( \frac{c}{\cos\theta} - c\right) NT.
\end{equation}

\begin{figure}
	\centering
	\includegraphics[width=\columnwidth]{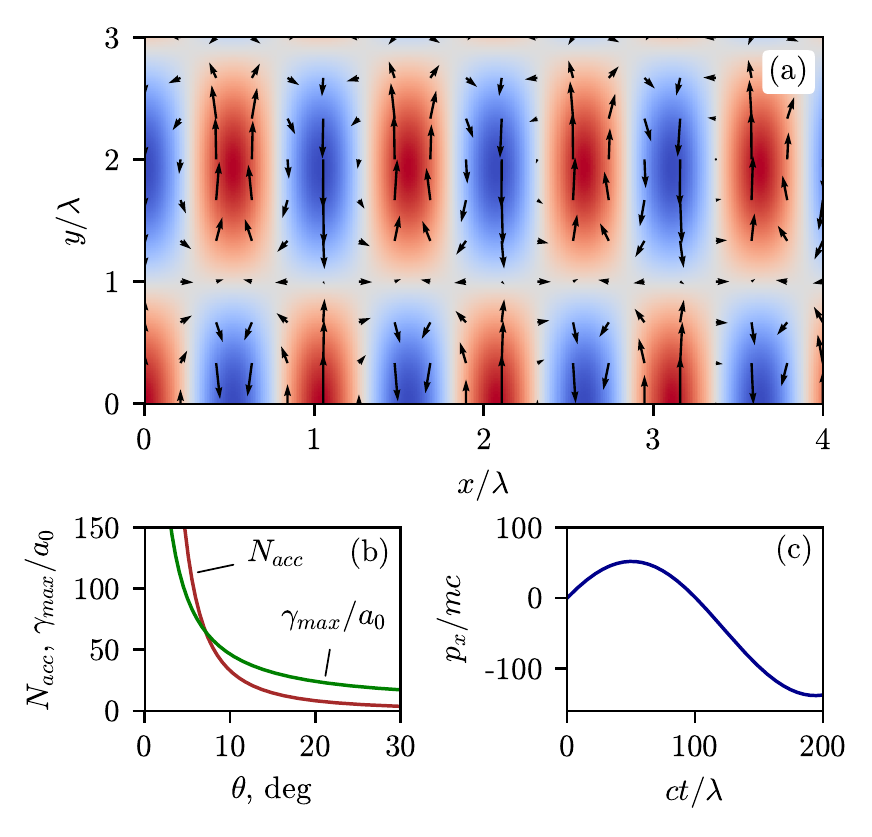}
	\caption{(a) Field structure near an ideally reflecting surface at $t=0$, $\theta = 15^\circ$. 
	Arrows depict the electric field ($E_x$ and $E_y$) relative value and direction, a color shows the value of the magnetic field $B_z$ (from blue to red). 
	The surface is located at $y=0$.
	(b) Functions $N_{acc}(\theta)$ and $\gamma_{max}(\theta) / a_0$.
	(c) Function $p_x(t)/mc$ for $a_0=5$, $x_0 = 0$ and $\theta = 12^\circ$.}
	\label{fig:N_gmax_px}
\end{figure}

The maximum possible displacement, which allows continuing acceleration, equals $\pi/2$. 
From this condition, one can determine $N_{acc}$, the maximum number of field periods during which the electron can be accelerated:

\begin{equation}
N_{acc}(\theta) = \frac{1}{2(1-\cos\theta)}.
\end{equation}
The function $N_{acc}(\theta)$ starts to grow very rapidly when $\theta$ approaches zero (see Fig.~\ref{fig:N_gmax_px}(b)); so a significant acceleration is possible only at small grazing angles and incidence angles close to $90^\circ$ (in so-called grazing incidence regime).

One can also calculate the maximum Lorentz factor that the electron can reach in such a process:
\begin{equation}
\label{eq:gamma_max1}
\gamma_{max}(\theta)-1 = \frac{e \overline{E_x} N_{acc} \lambda_x}{mc^2} \approx
\frac{2}{\pi}\frac{eE_0\cdot 2\sin\theta}{mc^2}\frac{N_{acc}\lambda}{\cos\theta}
\end{equation}
where $m$ and $e > 0$ are the electron mass and charge, respectively.
Here $\overline{E_x}$ is the averaged over time longitudinal electric field the electron feels in its own reference frame; since the electron speed $v_e\approx const=c$ except for a very short non-relativistic time period, $\overline{E_x} \approx 2E_0\sin\theta/\pi$.

So finally, we obtain the formula for the maximum electron Lorentz factor depending on the grazing angle:
\begin{equation}
\label{eq:gamma_max3}
\gamma_{max}(\theta) \approx 1 + \frac{4 a_0 \tan\theta}{1 - \cos\theta},
\end{equation}
whose plot is shown in Fig.~\ref{fig:N_gmax_px}(b). 
As well as $N_{acc}(\theta)$, the function $\gamma_{max}(\theta)$ is also growing as the grazing angle approaches zero.
So in the electron is placed exactly at the point where only $E_x$ field presents in the wave structure, it can be efficiently accelerated up to hundreds of MeVs by even moderate laser field intensities (with $a_0 \sim 1$). 

\subsection{Transverse stability}

The above calculations can be applied only to the electrons which are initially placed exactly at the nodes of the $E_y$ and $B_z$ fields. 
However, in the geometry described by fields (\ref{eq:oblique-fields}), these points form a set of measure zero; so to understand whether a significant fraction of electrons can be accelerated up to $\sim mc^2\gamma_{max}$ or only a negligible portion of them, one should consider test electrons with a small displacement from the node position ($y_0 = \Lambda/4 + \delta y, |\delta y| \ll \Lambda$) and calculate the maximum energy in that case.
If we switch to dimensionless variables $\hat{t}=\omega t$, $\hat{x}=2\pi x/\lambda$, $\hat{\mathbf{v}}=\mathbf{v}/c$, $\hat{\mathbf{p}}=\mathbf{p}/mc$, $\hat{\mathbf{E}}=e\mathbf{E}/(mc\omega)$, the equation of electron motion along the $y$-axis can be written as follows:

\begin{multline}
\label{eq:px1}
\frac{dp_x}{dt} = -E_x(x,y,t) = \\ 
= -2a_0 \sin\theta \sin\left[\sin\theta\left(\frac{\Lambda}{4} + \delta y\right)\right] 
\sin(x\cos\theta-t+\phi_0) \approx\\ 
\approx -2a_0\sin\theta \sin(x\cos\theta-t+\phi_0)
\end{multline}
because $\Lambda \sin\theta = 2\pi$ and $|\delta y| \sin\theta \ll 1$. Similarly,
\begin{multline}
\label{eq:py1}
\frac{dp_y}{dt} = -E_y(x,y,t) + v_x B_z(x,y,t)= \\
= 2a_0 (v_x-\cos\theta) \cos\left[\sin\theta\left(\frac{\Lambda}{4} + \delta y\right)\right]
\cos(x\cos\theta-t + \phi_0) \approx \\
\approx -2a_0(v_x-\cos\theta) \delta y \sin\theta \cos(x\cos\theta-t+\phi_0).
\end{multline}

We choose the initial phase $\phi_0 = -\pi/2$ so that at $t=0$ the force acting on the electron with $x = 0$ accelerates it in the direction of the wave propagation.
So the equations of motion finally are:
\begin{equation}
\label{eq:pxpy-final}
\begin{aligned}
\frac{dp_x}{dt} &= 2a_0\sin\theta \cos(x\cos\theta-t), \\
\frac{dp_y}{dt} &= -2a_0(v_x-\cos\theta) \delta y \sin\theta \sin(x\cos\theta-t).
\end{aligned}
\end{equation}

The main approximation is that the electron transverse motion is much slower than the longitudinal one:
\begin{equation}
\label{eq:fastslow}
\left|\frac{dp_y}{dt}\right| \ll \left|\frac{dp_x}{dt}\right|, \quad\left|p_y \right| \ll \left|p_x\right|,
\end{equation}
so that the motion can be split into 'fast' longitudinal and 'slow' transverse components.
For the longitudinal motion
\begin{equation}
\label{eq:px_int1}
p_x(t) = \int_0^t 2a_0\sin\theta \cos\left[x(t)\cos\theta - t\right].
\end{equation}

In the limit $a_0 \gg 1$, the electron becomes relativistic much quicker than the field period, and it can be shown that 
\begin{equation}
\label{eq:y_t}
x(t) \approx t+x_0-\frac{1}{2a_0\sin\theta \cos(x_0\cos\theta)} = t + x_0 - \Delta x_0,
\end{equation} 
for $t \gg \Delta x_0$. 
Here $x_0$ is the initial electron $x$-position.
For $\Delta x_0 \ll 1$, one should also assume that $|x_0\cos\theta| \lesssim \pi/4$.

Using Eq.~(\ref{eq:y_t}), we can also obtain the limitation on the initial positions $x_0$ and $\delta y_0$ for which the approach (\ref{eq:fastslow}) is applicable:
\begin{equation*}
|(v_x - \cos\theta) \delta y_0 \sin\theta \sin(x\cos\theta) | \ll |\sin\theta \cos(x\cos\theta) |,
\end{equation*}
\begin{equation}
|\tan(x_0 \cos\theta) \delta y_0 | \ll \frac{1}{|v_x - \cos\theta|} \le \frac{1}{\cos\theta},
\end{equation}
which is always satisfied as $|\delta y_0|$ is small.
Here we have also assumed that $t \ll (1-\cos\theta)^{-1}$ and $\Delta x_0 \ll 1$. 

Since the time interval [0;$\Delta x_0$] makes very a small contribution to the integral (\ref{eq:px_int1}), one can perform integration as if the electron is relativistic from the very beginning and write
\begin{equation*}
p_x(t) \approx 2a_0 \sin\theta \int_0^t\cos\left[(\cos\theta - 1)\,t + \left(x_0 - \Delta x_0\right)\cos\theta\right],
\end{equation*}
\begin{equation}
\label{eq:px_t_final}
p_x(t) \approx \frac{2a_0\sin\theta}{\Omega} \left[\sin\left(\Omega t - \psi_0\right) + \sin\psi_0\right],
\end{equation}
with the initial condition $\left.p_y\right|_{t=0} = 0$. 
Here, $\Omega = 1-\cos\theta$, and $\psi_0$ is determined by $x_0$, $a_0$ and $\theta$:
\begin{equation}
\label{eq:psi0}
\psi_0 = x_0\cos\theta - \frac{\cot\theta}{2a_0\cos(x_0\cos\theta)}.
\end{equation}
A typical plot of function (\ref{eq:px_t_final}) is shown in Fig.~\ref{fig:N_gmax_px}(c). 
The maximum of $p_x$ occurs at $t_{max} = \Omega^{-1}(\pi/2 - \psi_0)$ and the maximum value of $p_x$ described by (\ref{eq:px_t_final}) equals
\begin{equation}
p_x^{max} = 2 a_0 (1-\sin\psi_0) \frac{\sin\theta}{\Omega}
\end{equation}
which is of the order of estimate (\ref{eq:gamma_max3}).

If $t \ll \Omega^{-1}$, the momentum grows linearly:
\begin{equation}
p_x(t) \approx 2a_0 t\sin\theta \cos\psi_0.
\end{equation}

After the 'fast' motion components $p_x(t)$ and $v_x(t) \approx p_x (1+p_x^2)^{-1}$ are determined, we have to calculate the 'slow' motion from Eq.~(\ref{eq:py1}):
\begin{align}
\label{eq:dpx_dt}
\frac{dp_y}{dt} &= -2a_0 (v_x-\cos\theta) \delta y \sin\theta \sin(x\cos\theta - t) \\
\label{eq:ddeltax_dt}
\frac{d(\delta y)}{dt} &= \frac{p_y}{\sqrt{1 + p_x^2}}.
\end{align}

Substituting $x(t)$ in (\ref{eq:dpx_dt}) with (\ref{eq:y_t}) and supposing that $t \ll \Omega^{-1}$ (but still much greater than $1/\Delta x_0$), one may simplify the equations:
\begin{equation*}
\frac{dp_y}{dt} \approx -2a_0 \Omega \delta y \sin\theta \sin(-\Omega t + \psi_0),
\end{equation*}
\begin{equation}
\left\{
\begin{aligned}
&\frac{dp_y}{dt} = -2a_0 \Omega \delta y \sin\theta (\sin\psi_0 - \Omega t \cos\psi_0)\\ 
&\frac{d(\delta y)}{dt} = \frac{p_y}{2 a_0 t \sin\theta \cos\psi_0},
\end{aligned}
\right.
\end{equation}
or, finally,
\begin{equation}
\label{eq:deltay_t_full}
\frac{d^2(\delta y)}{dt^2} + \frac{1}{t} \frac{d(\delta y)}{dt} - \Omega^2 \delta y + \frac{\Omega\tan\psi_0 \delta y}{t} = 0.
\end{equation}

If we consider $t \ll \Omega^{-1}|\tan\psi_0|$, then the item $\Omega^2 \delta y$ may be omitted; then this equation can be reduced to the Bessel equation of $\sqrt{t}$ argument.
Its solution is:
\begin{equation}
\begin{aligned}
\delta y (t) = C_1 J_0(\sqrt{\Omega_1 t}) + C_2 Y_0(\sqrt{\Omega_1 t}),\quad \psi_0 > 0, \\
\delta y (t) = C_1 I_0(\sqrt{\Omega_1 t}) + C_2 K_0(\sqrt{\Omega_1 t}),\quad \psi_0 < 0,\\
\end{aligned}
\end{equation}
where $J_0$ and $Y_0$ are the Bessel functions of first and second kind, respectively; $I_0$ and $K_0$ are the modified Bessel functions; $\Omega_1 = \sqrt{4\Omega|\tan\psi_0|}$.
The initial conditions are: $\delta y(0) = \delta y_0$, $\delta y'(0) \approx 0$. Strictly speaking, $\delta y'(0)$ is not is not exactly zero because of non-relativistic motion at $t \lesssim \Delta x_0$, where $|E_y|$ is significantly greater than $|v_x B_z|$ and the electron gains some transverse velocity.
However, we suppose that $v_y \ll v_x < 1$, so $\delta y'(0) \ll 1$ always and $Y_0$ and $K_0$ with singularity at $t=0$ should be discarded. 
Therefore
\begin{equation}
\label{eq:deltay_t}
\begin{aligned}
\delta y (t) \approx \delta y_0 J_0(2\Omega_1 \sqrt{t}),\quad \psi_0 > 0, \\
\delta y (t) \approx \delta y_0 I_0(2\Omega_1 \sqrt{t}),\quad \psi_0 < 0,\\
\end{aligned}
\end{equation}
which means that for a relativistic particle, the equilibrium at $y = \Lambda/4$ is stable if $\psi_0 > 0$, and unstable otherwise. 
For typical parameters $(a_0 \gg 1, \theta \sim 5-15^\circ$) it means that the stability region is $x_0>0$.

At sufficiently large times ($t \gtrsim \Omega^{-1}$) $\delta y$ is not properly described by the above simplified equations.
However, at those times, the electron longitudinal force reaches its maximum and starts to decrease (see Fig.~\ref{fig:N_gmax_px}(c)) so those times are not of interest in the scope of our analysis.

If $\psi_0 \approx 0$, then the term $\Omega^2 \delta y$ in (\ref{eq:deltay_t_full}) cannot be omitted and Eq.~(\ref{eq:deltay_t}) is not correct.
For this case (and the other cases that do not satisfy the approximations of the current section), the equations of motion should be integrated numerically.

\section{Numerical model}
\label{sec:numerical-model}

\begin{figure*}
	\includegraphics[width=\textwidth]{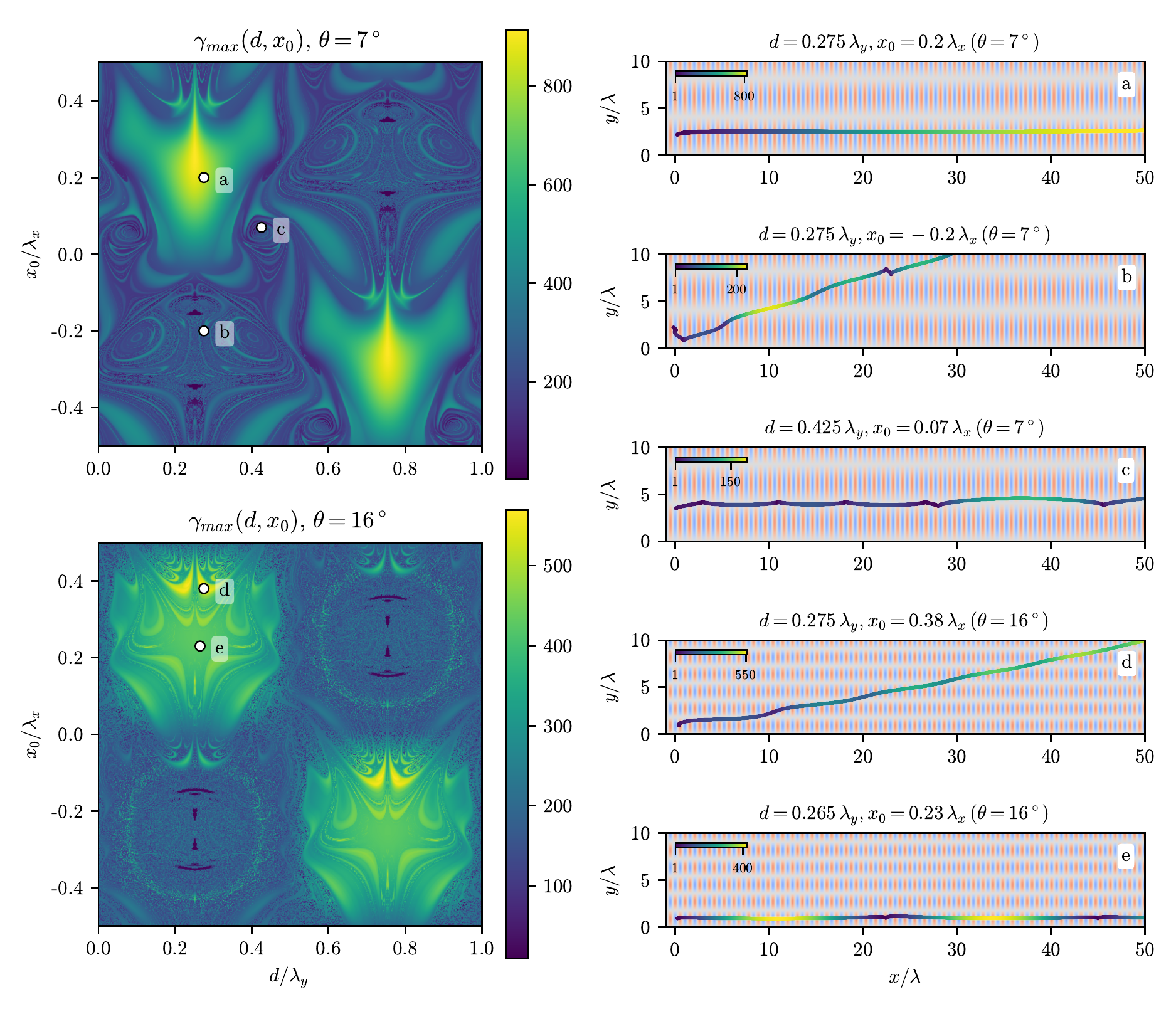}
	\caption{(Left) $\gamma_{max}(d, x_0)$ distribution at $a_0=16$ and $\theta = 7^\circ$ and $\theta = 16^\circ$.
	(Right) Sample electron trajectories for different initial positions (points (a)-(e) on the left plots). 
	The color denotes the value of the electron Lorentz factor over the trajectory,  
	the background color shows the $E_x$ field spatial distribution at $t=0$.}
	\label{fig:gmax_maps_traj}
\end{figure*}

In the previous section, multiple assumptions were made to estimate the maximum electron energy and the transverse stability condition. 
In the situations where those conditions are not satisfied, the equations of electron motion are not solved analytically and one needs to integrate them numerically. 
In the geometry where only $E_x$, $E_y$ and $B_z$ field components are present, $p_z$ is conserved and we can write the equations of motion of a test electron in the $xy$-plane as follows:

\begin{equation}
\label{eq:xy-motion}
\begin{aligned}
\frac{dp_x}{dt} &= - E_x - v_y B_z, \\
\frac{dp_y}{dt} &= - E_y + v_x B_z, \\
\frac{dx}{dt} &= \frac{p_x}{\gamma},\quad \frac{dy}{dt} = \frac{p_y}{\gamma},\\
\gamma &= \sqrt{1+p_x^2+p_y^2},
\end{aligned}
\end{equation}
Here $E$ and $B$ fields are taken from (\ref{eq:oblique-fields}).
We specify the following initial conditions: $x(0) \in [-\lambda_x/2, \lambda_x/2]$, $d \in [0, \lambda_y]$, $p_x(0) = p_y(0) = p_z(0) = 0$, where $d$ is the initial electron distance from the surface.
Since the fields are periodic (over $x$- and $y$-coordinates), the initial position can be taken within one spatial period without loss of generality.
In the model, we do not take into account that the fields cannot be described by (\ref{eq:oblique-fields}) below the surface (at $y<0$) and assume that the electrons always stay above the surface.

The equations are integrated up to $t_{max}$ that should be greater than $2\pi/\Omega$ (see Eq.~(\ref{eq:px_t_final})).
As the electrons experience the external field, their energy may increase or decrease with time; but the key parameter describing the acceleration is the maximum Lorentz factor of the electron $\gamma_{max}$. 
As one can see from Fig.~\ref{fig:N_gmax_px}(c), typically it is reached at the times of the order of $1/\Omega$, depending on the initial position. 
So if the laser pulse duration is properly selected and the field 'turns off' at around $t=\pi/(2\Omega)$, one can expect that the bunch of the accelerated electrons will have properties similar to that of the test electrons inside the field structure.

In Fig.~\ref{fig:gmax_maps_traj} (left), we present typical $\gamma_{max}$ distributions over the initial electron position. 
The parameters are: $a_0 = 16$, $\theta = 7^\circ$ and $\theta = 16^\circ$. 
As expected, the maximum possible acceleration can be achieved around $d = 0.25 \lambda_y$, for which the efficient acceleration occurs if $x_0 > 0$, and for $x_0 < 0$ the electrons typically do not gain much energy.
So this is in agreement with our theory~(\ref{eq:deltay_t}).

The maximum electron Lorentz factor for $\theta=7^\circ$ is about 900 which aligns with the estimate (\ref{eq:gamma_max3}). 
The important result of the modeling is that the acceleration is stable: high electron energy can be achieved in a relatively wide range of the $d$ and $x_0$.
For $\theta=7^\circ$, the stability regions are: $d/\lambda_y \approx 0.25\pm0.1$ and $x_0/\lambda_x \approx 0$--$0.3$, and the symmetrical region at $d/\lambda_y \approx 0.75\pm0.1$).

Several characteristic regions can be seen in the $\gamma_{max}$ distributions. 
Considering $\theta=7^\circ$ case (Fig.~\ref{fig:gmax_maps_traj}, left upper), we see, first, the region of the most efficient acceleration at around $d = 0.25 \lambda_y$, described in Sec.~\ref{sec:acc-model}. 
The typical electron trajectory for this region is shown in Fig.~\ref{fig:gmax_maps_traj}a: it is almost a straight line except for the non-relativistic part of the trajectory where the electric and magnetic parts of transverse Lorentz force do not compensate each other.
There are also chaotic regions where the electron is constantly 'jumping' between nodes and antinodes of the electromagnetic field (see Fig.~\ref{fig:gmax_maps_traj}b and Fig.~\ref{fig:gmax_maps_traj}c). 
The chaotic regions do not result in high electron energy because an electron does not feel the accelerating field for a long time there.

For a greater grazing angle $\theta=16^\circ$ (Fig.~\ref{fig:gmax_maps_traj}, left lower), the maximum electron Lorentz factor is about 570. 
It is significantly lower that in the case of $\theta = 7^\circ$. 
An interesting result is that the maximum $\gamma$ is achieved not at $d/y_0 = 0.25$ but at a bit different point.
If we look closely at the 'most energetic' trajectory (Fig.~\ref{fig:gmax_maps_traj}(d)) which corresponds to the point with the maximum Lorentz factor, we can see that the trajectory is different from what was considered in Sec.~\ref{sec:acc-model}.
Here the electron gains some energy in one semi-period of the standing wave along $y$-axis, and then, as soon as it is no longer in the proper phase of the field, it moves to the next semi-period where $E_x$ changes its sign and it again appears in the accelerating phase of the field.
In this case, it becomes possible for an electron to gain energy during a time interval greater than $1/\Omega$.
In contrast, the trajectory (e) (which is similar to (a)) corresponds to an electron which is accelerating only until $t\approx 1/\Omega$, then slowing down up to almost zero momentum and then accelerating again.
So in this case, the value of $\gamma_{max}$ is greater than predicted by Eq.~(\ref{eq:gamma_max3}).
All that means that for large $\theta$, the regime described in Sec.~\ref{sec:acc-model} does not result in the most efficient acceleration.

The dependence of $\gamma_{max}$ over $\theta$ and $a_0$ is depicted in Fig.~\ref{fig:gmax_a0_theta_spectra}(a, b), with detailed explanation in the next section.

\section{Particle-in-cell simulations}

The model considered above has several main limitations.
First, we assume that the wave incident onto a target is an infinite plane wave.
Second, we suppose that it is ideally reflected from the surface.
Also we assume that the electrons that subject to acceleration are essentially test electrons which do not alter the field structure.
These assumptions may be not satisfied in real world: the wave structure size is determined by the incident laser pulse which is usually focused on a small spot to achieve $a_0 > 1$; the target material (which becomes fully or partially ionized under the laser field) does not have infinite conductivity.
Also, it is usually desired to obtain high-charge electron bunches, and for that one needs to place large amount of electrons into the field so the field will differ from the idealized situation.
The problem of electron injection into the field also was not considered in the model and needs to be studied using different methods.

\begin{figure}
	\centering
	\includegraphics[width=\columnwidth]{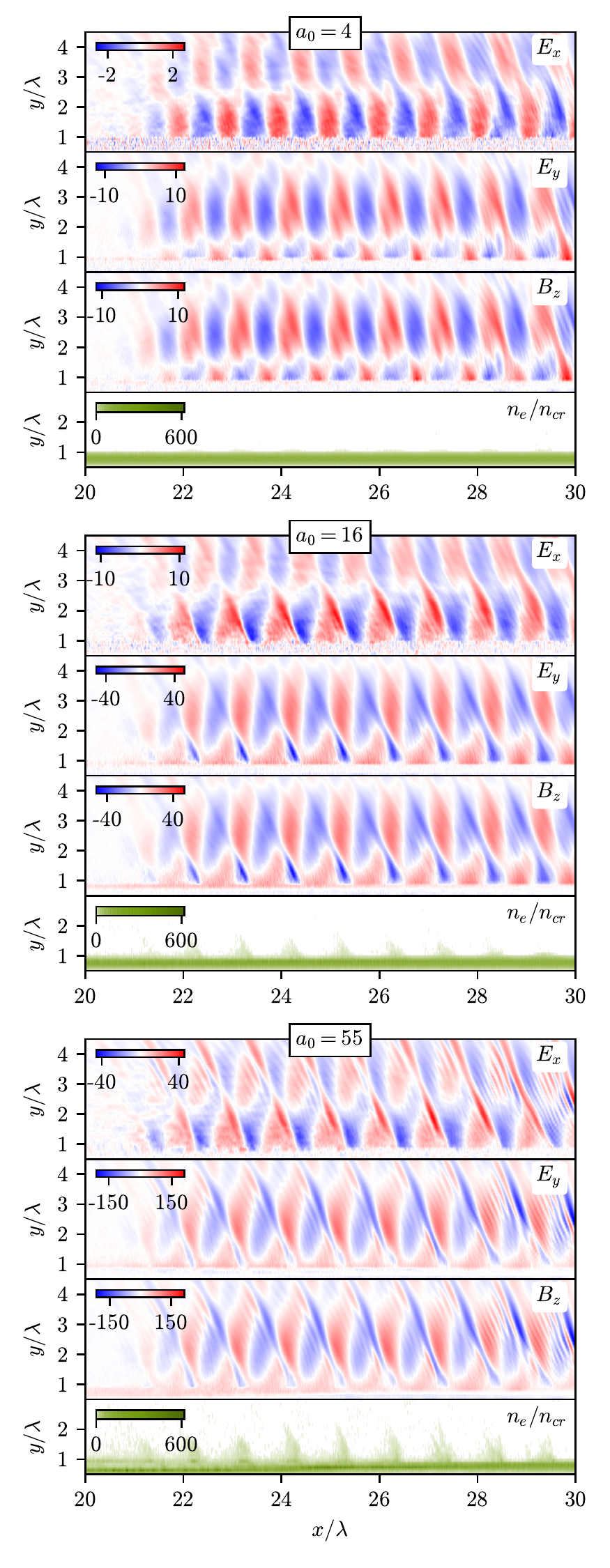}
	\caption{Longitudinal ($E_x$) and transverse ($E_y$) electric fields, magnetic field ($B_z$) and electron density ($n_e/n_{cr}$) in PIC simulations at different $a_0$ and $\theta = 12^\circ$.}
	\label{fig:ex_edensity_a0}
\end{figure}

Particle-in-cell simulations were performed to analyze the electron acceleration in more realistic situation. We used QED-PIC code \textsc{quill}~([\onlinecite{Nerush2010VANT}],  also see Sec.~II of [\onlinecite{Nerush14}] for details) which allowed us to evaluate the effect of radiation reaction on the acceleration process at higher $a_0$.
The simulations were performed for $a_0$ in the range of 4 to 55.
The laser pulse had $\lambda=1\mu m$, was $p$-polarized and had box-like shape (with flat longitudinal and transverse profiles with rapid decrease at boundaries). 
This shape was chosen to make the simulations closer to the model where the infinite plane wave is incident onto a target and to reduce computational complexity of the problem by making the domain smaller.
The laser pulse duration was 36 fs and the laser pulse width was 6 $\mu$m.
The grazing angle $\theta$ varied from $6$ to $18$ degrees.
The target was fully ionized with thickness of $0.5\lambda$ and electron density of $300 n_{cr}$ ($n_{cr} = \pi mc^2/(e^2 \lambda^2)$ is a critical plasma density) which corresponds to typical solid-state targets.
The charge-to-mass ratio of the ions was equal to 0.5 of that for a proton.
In the simulations, the step size was $dx\times dy \times dz = 0.008 \times 0.025 \times 0.1 \lambda^3$, the number of particles per cell was 9.

The simulations reveal that the field structure and target surface shape essentially depend on $a_0$.
In Fig.~\ref{fig:ex_edensity_a0}, spatial distribution of $E_x$, $E_y$, $B_z$ fields and the normalized electron density in the $xy$-plane are presented.
All values are taken at the same time $t=20\lambda/c$, the grazing angle was $12^\circ$.
At $a_0 = 4$ (upper plot), the field structure much resembles the case of superposition of two crossed plane waves (which was considered in the model).
Here $E_x \ne 0$ at the boundary due to the fact that the plasma conductivity is not infinite and a skin layer appears near the surface, but the modification of the boundary condition should only lead to a fixed field offset by $y$ and should not change the electron dynamics as the field is periodic.
At $y \approx y_{node} \approx 1.4 \lambda$, the transverse fields are almost zero, and longitudinal field is near its maximum, so the fields should allow efficient electron acceleration as described in Secs.~\ref{sec:acc-model} and \ref{sec:numerical-model}.
The plasma surface at $a_0=4$ appears to be almost unperturbed.

However, this is not the case for higher $a_0$.
As seen in Fig.~\ref{fig:ex_edensity_a0}, at $a_0 = 16$ and especially at $a_0=55$ a large portion of electrons becomes extracted from the target by the laser field, and the electron density in these areas is comparable to that of the target itself.
Therefore, the laser pulse no longer reflects ideally from a planar target, and the field becomes different from superposition of two monochromatic waves.
Essentially, generation of high-order harmonics in the reflected light is observed (Fig.~\ref{fig:ex_edensity_a0}, $a_0=55$). 
However, some key features of the field structure retain even at high $a_0$: the fields are still periodic along $x$- and $y$-axes and the phase speed is slightly superluminal.

\begin{figure}
	\centering
	\includegraphics[width=\columnwidth]{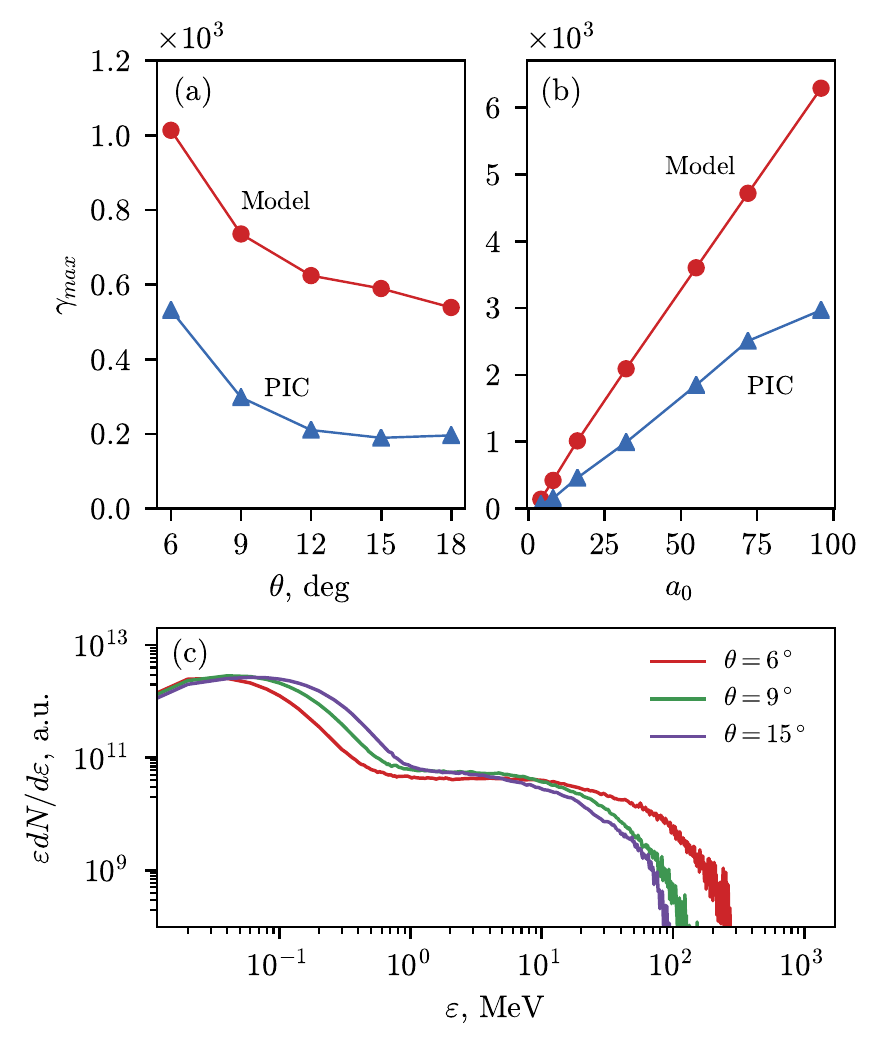}
	\caption{Maximum electron Lorentz factor $\gamma_{max}$ in the model and PIC (a) at different $\theta$ and $a_0=16$; (b) at different $a_0$ and $\theta=6^\circ$. (c) Electron spectra in PIC simulations at $a_0=16$ and $\theta=6$, $9$ and $15^\circ$.}
	\label{fig:gmax_a0_theta_spectra}
\end{figure}

When a $p$-polarized laser pulse is obliquely incident onto a plasma target, the surface electrons are pulled out of the target by the $E_y$ field [\onlinecite{Brunel87}].
These electrons may be pulled up to the point where the transverse field almost disappears.
As the longitudinal field is essentially non-zero there, the electron can be captured by the field (if it falls into the proper phase) and quickly accelerates up to high energies while moving in the $x$-direction.
This self-injection mechanism allows producing high-charge electron bunches.
In Fig.~\ref{fig:gmax_a0_theta_spectra}(c), the electron spectra from different simulations are depicted, while Fig.~\ref{fig:gmax_a0_theta_spectra}(a) and (b) show dependence of the maximum electron Lorentz factor on $a_0$ and $\theta$ parameters.
The maximum electron energies predicted by the model are also shown there.
Regarding the electron spectra, it may be split into two parts: the part where the electrons have energies of the order of $mc^2 a_0$ (due to stochastic heating by the laser field), and the energetic part with energies up to hundreds of MeVs (at $a_0=16$, see Fig.~\ref{fig:gmax_a0_theta_spectra}c).
With increasing $\theta$, the maximum electron energy decreases, and the trend is similar to that in the model (Fig.~\ref{fig:gmax_a0_theta_spectra}a), although the maximum energy in the simulations is about 2--3 times lower.
The linear scaling from $a_0$ is seen in PIC ($\gamma_{max} \approx 64 a_0$) as in the model ($\gamma_{max} \approx 35 a_{0}$), where the scalings are given for $\theta=6^\circ$.
However, $\gamma_{max}$ growth in PIC begins to saturate at $a_0 > 72$, which can be explained by the fact that plasmas well reflect the incident light only if $a_0$ is significantly less that the relativistically corrected critical density $n_{cr\,rel} = a_0 n_{cr}$ [\onlinecite{Gordienko05, Serebryakov15}], and in the simulations the plasma density was $300 \,n_{cr}$.
Also at such $a_0$, the radiation reaction force can also decrease the amount of energy the electrons can gain from the field.

\begin{figure}
	\centering
	\includegraphics[width=\columnwidth]{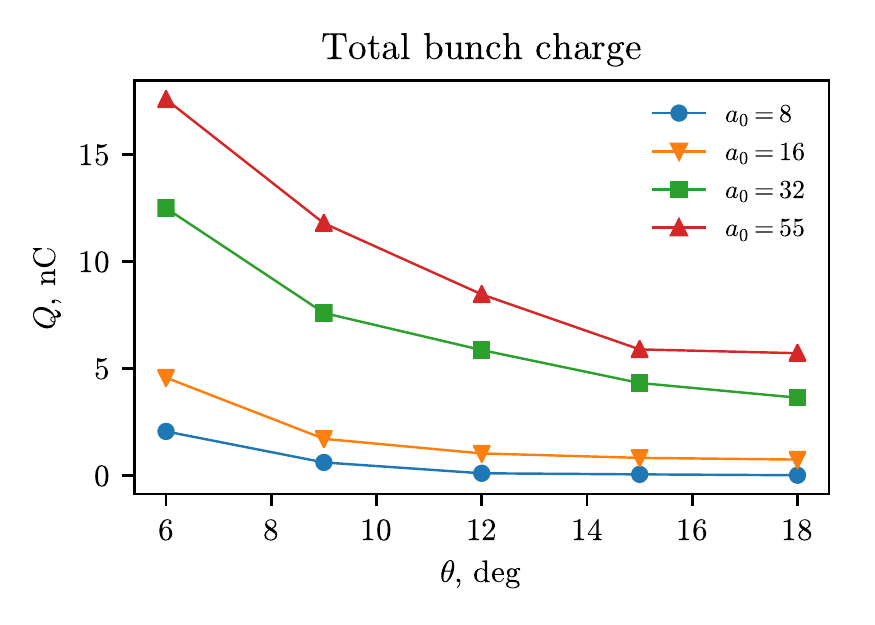}
	\caption{Total electron bunch charge in PIC simulations at different $a_0$ and $\theta$.}
	\label{fig:total_charge}
\end{figure}

An important feature of the considered acceleration scheme is very high (up to tens of nC) total electron bunch charge.
In Fig.~\ref{fig:total_charge}, the total bunch charge in PIC simulations at different $a_0$ and $\theta$ parameters is shown.
The bunch charge was calculated as the total charge of electrons with Lorentz factors $\gamma > 2 a_0$ and distance from the surface $d > 0.1 \lambda$.
As well as the maximum Lorentz factor, the bunch charge significantly increases as the grazing angle decreases.
The total charge equals a few nC (up to 2.32 nC) even at relatively low $a_0=8$.
At $a_0=16$ and $\theta=6^\circ$, the total charge was 4.57 nC, and at $a_0=55$ and $\theta =6^\circ$ it was equal to 17.5 nC.
Importantly, this is about 2--3 orders of magnitude more than a typical charge of an electron bunch produced by LWFA in gas targets, which looks very promising for many applications.

\begin{figure}
	\centering
	\includegraphics[width=\columnwidth]{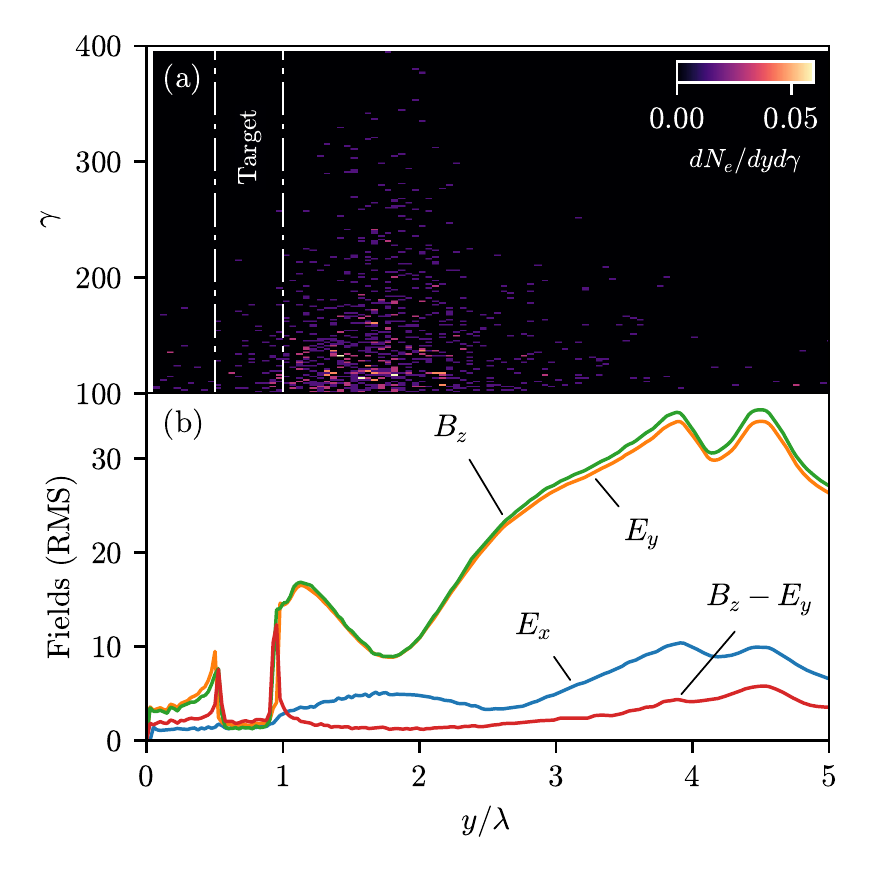}
	\caption{(a) Distribution of electrons with $\gamma > 100$ over $y$ and $\gamma$. 
	(b) Root mean square of fields ($E_x$, $E_y$, $B_z$, $B_z-E_y$, in $a_0$ units) at different $y$.
	Parameters: $t=24\lambda/c$, $a_0=32$, $\theta=15^\circ$.
	The target $y$-position is shown schematically in the upper figure.}
	\label{fig:edist_y}
\end{figure}

In order to understand the actual acceleration mechanism in the simulations, we have analyzed the distribution of the energetic electrons over the transverse coordinate.
In Fig.~\ref{fig:edist_y}(a), the distribution of electrons with the Lorentz factor greater than 100 is depicted, while in Fig.~\ref{fig:edist_y}(b) the average (root mean square) electric and magnetic fields are shown as a function of $y$.
Averaging was performed over $x$ in the region where both incident and reflected fields were interacting (e.g. from $x=24\lambda$ to $28\lambda$ in Fig.~\ref{fig:ex_edensity_a0}), and the fields were taken at $z$-coordinate corresponding to the center of the laser pulse.
The simulation parameters were: $a_0 = 32$, $\theta = 15^\circ$.
It is clearly seen that these electrons are concentrated in the area where the longitudinal electric field reaches its maximum value and the transverse electric and magnetic field reach their local minimums, so the total effective field acting on an ultrarelativistic electron ($B_z - E_y$ in the present system of units) is several times less than the longitudinal field.
That means that the mechanism described in Sec.~\ref{sec:acc-model} is actually realized, and accounts for almost all high-energy electrons in the simulations.

The significant difference of the electron maximum energy in the model and PIC (2--3 times) can be explained by several statements.
First, the electrons should be able to be accelerated by the field $E_x$ over a long enough distance $L_{acc} \approx 0.5\lambda/(1-\cos\theta)$.
However, in the simulations, this is usually not satisfied because of quick laser pulse defocusing (see discussions in Sec.~\ref{sec:conclusions}), especially for smaller $\theta$.
Another difference from the model is that even near $y = y_{node}$ (the point where the longitudinal field reaches the local maximum) the transverse force acting on a relativistic electron ($F_\perp \propto E_y - B_z$) is not exactly zero.
As seen in Fig.~\ref{fig:edist_y}(b), for $a_0 = 32$ and $\theta=15^\circ$ it is, on average, about 4 times less than the longitudinal field, so the transverse drift over large distances $L\gg\lambda$ becomes significant and the acceleration may stop.

In Fig.~\ref{fig:edist_y} (and in Fig.~\ref{fig:ex_edensity_a0} at $a_0=16$ and especially $a_0=55$) one can see the increase of the magnetic field $B_z$ with respect to the transverse electric field $E_y$; this effect can be explained by the theory in Ref.~[\onlinecite{Nakamura2004}]: at high incidence angles strong surface electron currents generating surface-bound magnetic field appear on the plasma surface.
The presence of strong surface currents was also observed in our simulations.
However, this quasi-static magnetic field presents only at small distances from the surface and does not significantly affect the dynamics of the vacuum electrons (with $d\sim\lambda$).

\section{Preplasma as an electron injector}

One of the main advantages of the configuration of 'grazing incidence' is that there is no need for external electron injector as the electrons are being extracted from the target in a natural way (as in Refs.~\onlinecite{Brunel87, Naumova2004}).
However, it turns out that the presence of a very underdense plasma near the surface (for example, a preplasma formed by a laser prepulse) may further improve the electron acceleration efficiency.
Effect of preplasma on the electron dynamics near the target surface has been extensively studied over the last years~[\onlinecite{Ovchinnikov13, Ivanov17, SerebryakovQE17}].
As more electrons are present in the strong field region, more of them are likely to have the proper initial condition, and there is higher chance for any given electron to experience the maximum or near-maximum possible acceleration. 
Therefore, both the number of multi-MeV electrons and the maximum energy would be expectedly higher.
Nevertheless, at a certain density the plasma fields will be strong enough so that the resulting field structure will be too different and the accelerating electric field will become weaker.
Therefore, there should be the optimum preplasma density.

\begin{figure}
	\centering
	\includegraphics[width=\columnwidth]{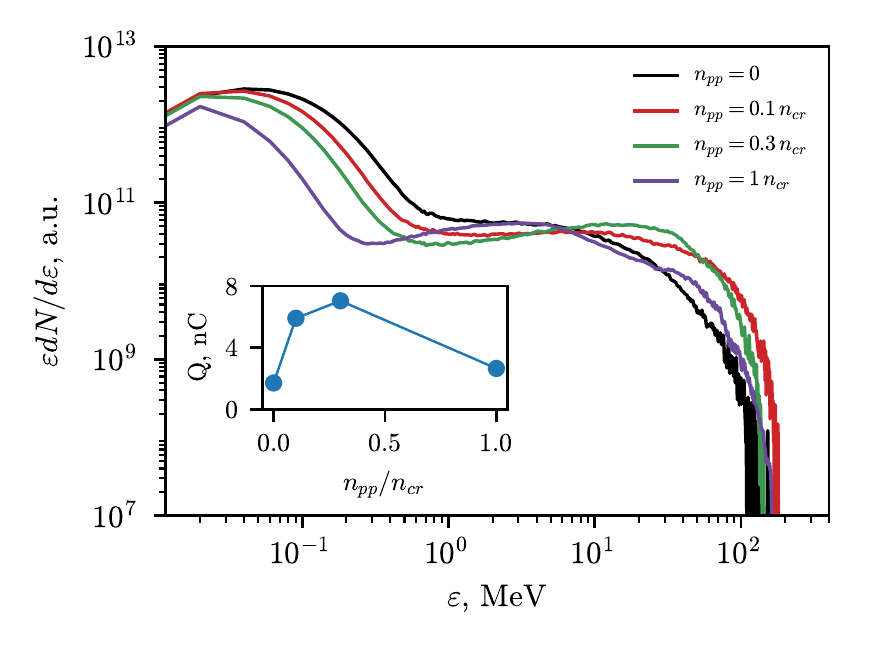}
	\caption{Electron spectra at the same parameters as on Fig.~\ref{fig:gmax_a0_theta_spectra}(c) ($\theta=9^\circ$) and different preplasma density $n_{pp}$.
	The inner plot shows the total bunch charge dependence on the preplasma density.}
	\label{fig:spect-pp}
\end{figure}

Our simulations confirm that the preplasma can actually improve the electron bunch properties. 
In Fig.~\ref{fig:spect-pp}, the black line (target without a preplasma) shows a relatively small number of high-energy electrons and a very large number of low-energy electrons. 
At preplasma densities $n_{pp}=0.1 n_{cr}$ and $n_{pp}=0.3 n_{cr}$, the number of electrons with energy $> 10$~MeV rises about $\sim2$--5 times, and the maximum energy is also slightly higher.
Increasing the preplasma density to $1 n_{cr}$ does not seem to improve the acceleration further; in contrast, the maximum electron energy starts to decrease.
It should be noted that the maximum density of a preplasma in experiments is typically about $1\,n_{cr}$~[\onlinecite{LePape2009}].
Also the total bunch charge is presented in the inner plot in Fig.~\ref{fig:spect-pp}; there is also an optimal value $n_{pp}=0.3 n_{cr}$ which results in the maximum bunch charge of 7.04 nC, which is about 4 times greater than without a preplasma.

In these simulations, most of the parameters were the same as in the previous section. 
Preplasma was simulated as a plasma layer with thickness $L_{pp}=2\lambda$ and consisting of the same matter as the target, with electron density decreasing linearly from $n_{pp}$ to 0. 
Real preplasmas usually have density profile close to exponential with typical scale length of the order of few microns~[\onlinecite{LePape2009}], but modeling of the exponential profile requires more computational resources, and using a linear profile may be considered as a first-order approximation.

\section{Conclusions and discussions}
\label{sec:conclusions}

In the current work, a model for electron acceleration by grazingly incident $p$-polarized laser pulses was developed and used to demonstrate the stability of electron acceleration at different incidence angles.
The model allowed to analyze the electron trajectories and it was shown that there are 2 different types of trajectories that result in the most efficient acceleration: almost longitudinal motion (which results in the maximum energy gain at lower grazing angles, $\theta=7^\circ$) and quasi-periodic step-like trajectory which turns to be the most efficient at higher grazing angles ($\theta=16^\circ$).
However, the total acceleration rate at higher grazing angles is much lower.

A set of particle-in-cell simulations was carried out to analyze the applicability of the results in more realistic situations.
It was shown that although the maximum electron energy in PIC is about 2--3 times lower at the same parameters, similar dependences on $\theta$ and $a_0$ are observed.
The maximum electron energy in PIC approximately scaled as $\gamma_{max}\approx 35 a_0$ so at $a_0=55$, the electrons with energy about 1 GeV were observed in the simulations.
The electron bunches had very high total charge (up to 17.5 nC at $a_0=55$).
Both the acceleration rate and the total charge showed similar dependencies on $\theta$, with the smallest considered angle $\theta=6^\circ$ resulting in the highest energy and bunch charge.
Also, the electron distribution over the transverse coordinate showed that energetic electrons in PIC were concentrated near the longitudinal field maximum and the transverse fields minimum, which is an evidence to that the acceleration mechanism considered in the model is actually realized. 
Also it was demonstrated that in the presence of a low-density preplasma near the target, the maximum energy of the electrons and the total bunch charge may increase.
There is an optimal preplasma density $n_{pp\, opt}\sim 0.1-0.3 n_{cr}$ so that at $n_{pp} > n_{pp\,opt}$ the acceleration rate and bunch charge become lower.

In the model, we did not take into account the laser pulse defocusing.
However, this effect may limit the maximum electron energy in the case of realistic laser pulses, because in experiments with laser-solid interaction, the laser pulses are usually highly focused on the surface so the light intensity at the target is maximized.
However, a tightly focused laser beam is subject to a rapid defocusing and is not well suitable for forming the field structures like in Eqs.~(\ref{eq:oblique-fields}).
One may estimate the characteristic laser pulse width that is required to form the field structure.
The typical distance at which the beam diverges is the Rayleigh length $r_d=\pi w_0^2/\lambda$, where $w_0$ is the laser pulse waist radius.
For an estimate, we can take the distance $0.5 r_d$ as the maximum distance at which the wavefront still roughly resembles the plane wave.
Also the distance needed for optimal electron acceleration equals $L_{acc} = 0.5\lambda/(1-\cos\theta)$. 
From the condition $0.5 r_d > L_{acc}$ one may estimate the optimal laser pulse waist diameter for a given $\theta$: $d_{opt} \approx 2 \lambda / \sqrt{\pi(1-\cos\theta)}$.
For example, $d_{opt} = 15.2\lambda$ at $\theta=6^\circ$, and $d_{opt} = 6.0\lambda$ at $\theta=15^\circ$.
It should be noted that focusing the laser pulse at a spot less than $d_{opt}$ in size should decrease the maximum electron energy, because in this case the acceleration distance $L_{acc}$ becomes determined by defocusing distance and not by $\theta$. 
The laser field intensity at the focus scales as $w_0^{-2}$ so the electric field scales as $w_0^{-1}$, while $r_d$ is proportional to $w_0^2$.
Therefore $\gamma_{max} \sim w_0$ so focusing of the laser pulse on a spot less than $d_{opt}$ in size is not optimal.

In a number of papers [\onlinecite{Li2006}, \onlinecite{Chen2006}], the near-surface electron acceleration is attributed to strong quasi-static electromagnetic fields that, along with the reflected pulse, allow the electrons to be in a resonance with the field and to gain energy.
However, under the conditions that occur in our simulations (interaction with a highly overdense plasma), the incident and reflected laser fields mostly determine the electron motion in the vacuum region near the target, while the plasma fields in this region do not rise strongly and almost do not affect the electron dynamics.
It could be shown as follows.
The quasi-static electric field produced by the electrons extracted from the surface is determined by the characteristic density of the vacuum electrons in the region $y \gtrsim y_{node}$ (if we consider acceleration in the vicinity of the first node of the transverse fields, or at $y\approx y_{node}\approx1.4$--1.6 in Fig.~\ref{fig:ex_edensity_a0}).
If we denote the characteristic density of the plasma above surface as $n_p$, then the ratio between the typical plasma field and the laser field can be estimated as follows:
$E_p/E_0 = mc\omega_p/(eE_0) = \omega_p/(\omega_L a_0) = a_0^{-1} \sqrt{n_p/n_{cr}}$.
For example, at $a_0=16$, $n_p\sim5 n_{cr}$ at $y\approx y_{node}$ in our simulations so the quasi-static electric field (which is transverse with respect to the electron motion) can be estimated as $0.13 a_0$.
At the same time, the transverse field from the laser is $\approx 2a_0$ so the electron motion is determined mostly by the laser field.

A very high total charge (of the order of 10 nC) of accelerated electrons is a characteristic feature of the considered source of energetic electrons, which appears because of relatively high electron density of solid-state targets.
Such charge is cannot be currently obtained with LWFA-based accelerators, where a typical charge of the electron bunches is usually of the order of tens or, in best cases, a few hundreds of pC~[\onlinecite{Esarey09, Malka-CERNYR211}].
However, according to our simulations, the electron bunches with total charge of 17.5~nC can be generated at the laser intensity about $10^{21}-10^{22}$ W/cm${}^2$ which is now available in many laboratories in the world.
So such sources of accelerated electrons based on laser-solid interactions look very promising for many applications, such as employing them as injectors in multi-stage accelerators, or producing ultrahigh-brilliance femtosecond gamma-ray beams.





\section{Acknowledgements}

The authors are thankful to the Russian Science Foundation (grant 16-12-10383) for financial support.

\bibliography{grazing-article}
\end{document}